\documentstyle[aps,preprint,epsf,psfig]{revtex}
\tighten
\draft

\begin{document}

\title{Neutron production by cosmic-ray muons at shallow depth \\}
\author { F. Boehm$^3$, J. Busenitz$^1$, B. Cook$^3$,
G. Gratta$^4$, H. Henrikson$^3$, J. Kornis$^1$ \\ 
D. Lawrence$^2$, K.B. Lee$^3$,  
K. McKinny$^1$, L. Miller$^4$, V. Novikov, A. Piepke$^{1,3}$, \\
B. Ritchie$^2$, D. Tracy$^4$, 
P. Vogel$^3$, Y-F. Wang$^4$, J. Wolf$^1$  \\
(Palo Verde Collaboration)}
\address{$^1$ Department of Physics and Astronomy, University of Alabama,
Tuscaloosa AL 35487 \\
$^2$ Department of Physics and Astronomy, Arizona State University,
Tempe AZ 85287 \\ 
$^3$Department of Physics, California Institute of Technology,
         Pasadena, CA 91125, USA \\
$^4$ Physics Department, Stanford University, Stanford CA 94305 }
\date{\today}
\maketitle

\begin{abstract}
The yield of neutrons produced by cosmic ray muons at a
shallow depth of 32 meters of water equivalent has been measured.
The Palo Verde neutrino detector, containing 11.3 tons of Gd
loaded liquid scintillator and 3.5 tons of acrylic served as a target.
The rate of one and two neutron captures was determined.
Modeling the neutron capture efficiency allowed us to deduce the
total yield of neutrons $ Y_{tot} = (3.60 \pm 0.09 \pm 0.31 )
\times 10^{-5}$ neutrons per muon and g/cm$^2$.
This yield is consistent with previous measurements at similar
depths.

\end{abstract}

\vspace{1cm}

\section{Introduction}

In the present work we report the results of a new
measurement of the muon-induced neutron production
at shallow depth, based on the Palo Verde
neutrino detector \cite{PV}.
Even though the device was designed for a different
purpose, namely the detection of reactor $\bar{\nu}_e$,
it was possible to operate it in a mode suitable for the identification
of muon-induced neutrons.

Neutrons and other hadrons produced by cosmic-ray muons in the earth are
an important and unavoidable source of background for underground low
rate experiments. Knowledge of the rates for these processes
is an essential part of understanding backgrounds in neutrino detectors
and other low counting rate experiments. For example, since muon-produced
fast single or multiple neutrons can mimic the correlated signature of
inverse neutron beta decay, searches for neutrino oscillations at nuclear
reactors \cite{PV,Chooz} must cope with this source of background. Other neutrino
and proton decay experiments, as well as dark matter searches (even
though often at greater depth), have to cope with this source of background as
well. The CDMS experiment, for instance, is searching for cold dark
matter \cite{Blas}, and is presently at shallow depth; muon-induced neutrons
represent a major source of background. Low-energy accelerator neutrino
oscillation searches \cite{LSND,Karmen2} 
are usually performed near the Earth's surface
where muon-induced neutrons are a significant source of background.
Another example is the proposed ORLaND neutrino detector at the
Spallation Neutron Source at Oak Ridge National Laboratory \cite{Orland} 
which is also planned to be at shallow depth.

Despite the importance of the subject, relatively little recent progress
has been reported. Several measurements of the neutron production rates
at various depths conducted in the past suggest a smooth dependence of
the neutron yield on depth or, equivalently, on the average muon energy
\cite{Bezrukov,Enikeev,Aglietta}. 
However, a more recent measurement by the LVD collaboration \cite{Aglietta2},
resulting in a smaller yield, is in disagreement with this simple
dependence on depth. At shallow depth, where the average muon energy is
$\sim$ 10 GeV, the smooth trend was confirmed by measurements with
relatively small detectors without much shielding against the neutrons
produced outside them \cite{Bezrukov,friends}. 
In the more recent experiment  \cite{friends}, the
single and double neutron yields were determined separately, and pion
production by muons was also observed. One drawback of that experiment,
possibly present in the other ones as well, is the difficulty of
distinguishing between the neutrons produced in the detector (the
intended source of neutrons) and the neutrons produced outside
by the muon-induced showers. In fact,
in Ref. \cite{friends} 
it has been estimated that half of the detected single
neutrons originated in the hadron showers outside the detector volume.

The theoretical description of this background process is usually based
on the assumption that the electromagnetic interaction of high-energy
muons with matter can be modeled by replacing the exchanged virtual
photon by ``equivalent" real photons, and using known photo-nuclear
reaction cross sections \cite{WW}. 
The analysis is complicated, since in order
to relate the theoretical neutron production yield to measurement, the
propagation and possible cascade multiplication of all reaction products
must be understood. (For example, a $\pi^-$ produced by a muon will make
more neutrons, as does the $(n,2n)$
reaction, etc.). While the smooth variation noted above for the neutron
reaction rate versus depth is supported in some calculations  \cite{Olga}, 
other approaches come to different conclusions\cite{Alk,Delorme}. 
In particular, Ref. \cite{Delorme} was devoted, 
unlike the others, to the relatively shallow depth
relevant for the present work. In that work, the calculated neutron yield
was smaller than the measured one  \cite{friends}, 
while the calculated and measured
$\pi^ +$ yields agreed with each other. One should also keep in mind that
while the theoretical description quoted above deals with the
muon-nucleus interaction involving the exchange of a virtual photon,
neutrons can be also produced by nuclear interactions involving real
bremsstrahlung photons, and electron/positron pairs created during the
passage of muons through matter.(The present experiment cannot 
separate neutrons created by
interactions involving virtual photons from those produced by
interactions of real photons.)

\section{Experiment}

The measurement of the neutron production rate was performed
at $32\pm3$ meter-water-equivalent (mwe) depth using the Palo Verde neutrino
detector. The detector and its operation is described in detail 
in Ref. \cite{PV}. Briefly, the apparatus consists of 66 acrylic
cells, each 900 x 12.7 x 25.4 cm$^3$. These cells are filled with liquid
scintillator loaded with Gd 0.1\% by weight, for a total scintillator
volume of 11.34 tons. The acrylic material has an aggregate mass of 3.48
tons. Muons can thus spall either in the scintillator volume or in the
acrylic material, and the resulting neutrons cannot be distinguished by
source.
Neutrons, when moderated to thermal energies, are
preferentially captured on Gd, resulting in a 
$\sim 8$ MeV gamma-ray cascade, the characteristic
neutron capture signal. The
central active detector volume is surrounded by a 1 m thick water shield,
and the outermost layer of the detector is an active muon veto counter
providing 4$\pi$ coverage.

Since the apparatus was intended for the
detection of reactor neutrinos, it was not optimized for the
neutron yield measurement.
For the present purpose special runs were performed without
the trigger rejection of the cosmic rays.
The throughgoing muon sample was selected off-line such that 
at least two veto hits 
were recorded. Only muons which at the same time went
through at least three cells of the central detector 
were included. The delayed neutron capture events,
recognized by their energy deposit, are of two kinds:
The ``single bank'' events, where only one neutron capture candidate
event occurs following the veto hit; and 
the ``two bank'' events, which have
two neutron capture candidate events following the muon.
(The detector electronics is not capable of recording 
more than two correlated events.)

The measured quantities are the numbers of single ($N_1$)
and double neutron-capture events ($N_2$) associated with $N_{\mu}$ muons
traversing the central detector. The average path length
of these muons in the central detector is $X$ (measured in units
of  g/cm$^2$). Since some of the neutrons could have been created in
the water shield or in other external detector parts,  correction factors 
$Q_k < 1$ are applied to $N_k,~k=1,2$. Finally, the neutron
detection efficiencies $\epsilon_{k,l}$ are introduced. Here,
e.g, $\epsilon_{1,1}$ is the probability that one neutron
was created and one detected, while $\epsilon_{1,2}$ is
the probability that two neutrons were created and only one detected.
If all these quantities were known
one could define  neutron
yields $Y_l$, etc. (per muon and  g/cm$^2$), 
where $l$ is the number of neutrons, 
that are independent of
the detector properties and obey the relation
\begin{equation}
N_k = \frac{N_{\mu} \cdot X \cdot \epsilon_k^{daq}}{Q_k} 
\cdot \sum_{l=1}^3 Y_l \cdot \epsilon_{k,l}  ~.
\end{equation}
Here $ \epsilon_k^{daq}$ is the detection
livetime correction which
in this case depends on the number of detected neutron-like
events $k$.

Obviously, since only two quantities, $N_1$ and $N_2$,
are measured, only two yields can be determined, and 
the system above cannot be
solved without  approximations.
It is assumed further that the contributions from four
or more produced neutrons can be neglected,
as indicated in equation (1). 
Moreover, 
as will be shown below, the $total$ neutron yield
\begin{equation}
Y_{tot} = Y_1 + 2 \cdot Y_2 + 3 \cdot Y_3
\end{equation}
is essentially independent of the ratio $Y_3/Y_2$,
i.e., on the assumed value of $Y_3$,
while the deduced single and double neutron yields
$Y_1$ and $Y_2$ depend on that ratio significantly.
Thus, the final results of the present experiment
will be expressed as the measurement of $Y_{tot}$.

Note that the quantities
$Y_k$, and thus also $Y_{tot}$, 
contain $all$ processes that lead to the production
of $k$ neutrons by the muon. Thus, if the muon creates a $\pi^-$
(or any other particle except a neutron) which then, in turn,
creates $k$ neutrons, all of the
neutrons, regardless of source, contribute to $Y_k$.
This somewhat awkward
definition is necessary since the evaluation of the efficiencies
$\epsilon_{k,l}$ is based on a code that tracks just neutron propagation
in the detector. 

The neutron capture events were recorded in two runs, each
about half of a  day long. The raw muon veto rate was 
about 2 kHz and the rate of muons
which went through at least three cells of the central detector
and caused two or more detectable veto hits in the first (second) run
was 270(275) Hz. These two runs are essentially equivalent, and
the final total number of muons was
determined to be $N_{\mu} = 1.42\times10^7$, very
similar in both runs.

The average path length $X$ of the muons is estimated
with a simple
ray-tracing Monte Carlo simulation,
starting from a  $\cos^2\theta$
zenith angle distribution.  The resulting
$X$ was 125 g/cm$^2$ for the central detector (i.e., the
scintillator, the acrylic cells and the small amount of other
materials (Cu, Fe) in the central detector) and 317 and 62 g/cm$^2$ 
for the water shield and veto counter, respectively.

The neutron capture events $N_k$ were selected using 
cuts similar to the neutron part of the neutrino signal \cite{PV}. 
For the single neutron events a time cut of 10 - 100 $\mu$s
between the last muon event and the neutron 
capture event was used, while for the two neutron events
only the lower limit of 10 $\mu$s was kept,
the upper one being irrelevant. The time
between the two neutron capture events was restricted to 5 - 100 $\mu$s.
The energy distributions of all neutron events had identical
shapes within statistics.  

There are several contributions to the livetime correction factors
$ \epsilon_k^{daq}$. The largest one, which affects primarily
the single neutron events, arises since the initial
information is overwritten when  another muon strikes the veto
after the single neutron event was recorded and before the
450 $\mu$s time interval of accepting new events.
Combining the correction for this effect with other deadtime
corrections, the resulting $ \epsilon_1^{daq}$ = 30.8(27.8)\%
for the one-neutron events in the first(second) runs, and
$ \epsilon_2^{daq}$ = 72.4(70.3)\% for the two-neutron
events. The difference in $ \epsilon_k^{daq}$ between the two
runs is caused primarily by the $\sim 8\%$ change in the
raw muon rate. 

The  distribution of the
time interval between the muon and the single-neutron
event is shown in the top panel of Fig. 1, while the lower
panel shows the same distribution for the first of the
two neutron events. The fitted curves are of the form
\begin{equation}
F(t) = (a_1 e^{-t/\tau_1} + a_2 e^{-t/\tau_2}
+a_3) \cdot e^{-t/500\mu{\rm s}} ~.
\end{equation}
Here $\tau_1 = 28.8 \pm 1.0 ~\mu$s for the single
neutron case while $\tau_1 =  11.8 \pm 0.4 ~\mu$s 
(not very far from  half of the
previous  $\tau_1$, the value one expects for two neutrons) 
for the double neutron
case, are the characteristic neutron capture times. Both 
agree quite well with the 
Monte Carlo (MC) simulation, as seen in Fig. 1.
The presence of the time constant  $\tau_2$
is related to the inhomogeneous nature of the detector.
Some neutrons enter the acrylic, where there is no Gd
available for captures, thus prolonging their capture time. 
Finally,
the last term, with the fixed time constant of 500 $\mu$s,
represents the accidental background (500 $\mu$s is the
average time interval between successive muons).
Using such a fit one can subtract the accidental background 
in the relevant time window, and calculate the corresponding 
correction for the described choice of the time cuts. 
  
In Fig. 2 the distribution of 
the elapsed time between the two neutron
capture events is shown, and again fitted to the same functional
dependence, Eq. (3). (The distinction between the
top and bottom panels is explained below.) 
Here, the time constant $\tau_1 = 18.1 \pm 1.0~ \mu$s
in the top panel
is substantially smaller than the $\tau_1$ obtained by the
MC simulation ($\tau_1^{MC} \sim 29 ~\mu$s, 
as expected if only two neutrons are involved).
This feature suggests that the data sample contains a 
nonnegligible component with more than two neutrons. 
We will return to this point later.

\section{Efficiencies}

While the evaluation of $N_{\mu}$, $N_1, N_2$, 
and $X$ is rather straightforward,
the determination of the efficiency matrix $\epsilon_{k,l}$ is 
somewhat model dependent. 
This is because, in order to calculate $\epsilon_{k,l}$, one
needs the initial energy and angular distributions of the created
neutrons, both of which are poorly known. In practice, a number of
physically plausible assumptions are made about these distributions, and
the spread among the resulting efficiencies yields a measure
of the systematic uncertainty.
Once the neutron initial distribution is chosen, the
Monte Carlo simulation of the neutron transport 
developed for the neutrino experiment \cite{PV} is used. 
The detector geometry, materials and electromagnetic
interactions are simulated using GEANT \cite{Geant}.
Hadronic interactions are simulated by FLUKA \cite{Fluka},
and the low energy neutron transport by GCALOR \cite{Gcalor}. 

To evaluate the quantities $\epsilon_{1,1}$ and $\epsilon_{2,1}$
initial single neutrons were  distributed randomly in position
and initial angles through the detector volume.
For the initial energy distribution,
several possibilities were used: the exponential
distribution $\exp^{-E/39{\rm MeV}}$
as proposed by the Karmen collaboration \cite{Karmen},
and functions with the power dependence 
$E^{-x}, ~~0.5 \le x \le 2$ (see \cite{Barton}). 
These cover the shape following 
$E^{-1.86}$ suggested by the
experience with photo-nuclear processes \cite{Khalchukov}. 
All of these distributions result in 
similar efficiencies. Using the spread of the calculated
values as a measure of the systematic error, and taking
a simple average as the most probable value, one obtains
\begin{equation}
\epsilon_{1,1} = 0.20 \pm 0.03, ~~ \epsilon_{2,1} = 0.007 \pm 0.004 ~.
\end{equation}
Note that the probability that a single initial neutron will
result in two neutron captures, characterized by  $\epsilon_{2,1}$,
is quite small.

The determination of two neutron efficiencies is even more difficult,
since it depends on the energy distribution of both initial neutrons.
To evaluate the quantities $ \epsilon_{1,2}$ and $\epsilon_{2,2}$
two neutrons were created at the same random point and with  random 
directions for each of them. Two extremes were considered.
In the first of them, the total energy of the neutrons followed 
one of the previously described
functions used above in the single neutron case. This initial
energy was distributed randomly between them. (Essentially
the same efficiency was obtained if the two neutrons shared the
initial available energy equally.)
The other extreme is obtained if both neutrons  each have the energy
distribution used in the single neutron case, i.e., the total
neutron energy is on average twice as large. The spread
(not very large) was again used in the averaging and in the
estimate of the systematic error, resulting in
\begin{equation}
\epsilon_{1,2} = 0.22 \pm 0.01, ~~ \epsilon_{2,2} = 0.06 \pm 0.01 ~.
\end{equation}

Note that there is a sizable probability that only one neutron
is detected when two were initially created. None of the 
previous analyses, in particular Ref. \cite{friends}, took that
into account. Next, one has to consider the possibility that
three neutrons were spalled by the muon, but only one or
two were detected. Again, using the average and the two
extreme possibilities to divide the available energy among
the three neutrons, one obtains
\begin{equation}
\epsilon_{1,3} = 0.19 \pm 0.01, ~~ \epsilon_{2,3} = 0.10 \pm 0.01 ~.
\end{equation}
It is important to realize that these efficiencies are not
much  smaller than those for initial one or two neutrons.
Thus, the effect of $Y_3$ should be considered. 
(The effect of four and more neutron spallation will be neglected,
however.)

The correction factors $Q_k$ that exclude neutrons created outside
the central detector volume
(i.e. in the water shield since the effect of the veto is negligible),
but captured there, must be determined also.
To do that,
the efficiencies for each passive volume were determined using 
the same Monte Carlo code as for the case of the central detector.
Then $Q_k$ were determined from
\begin{equation}
Q_k = \frac{(X \sum_l Y_l \epsilon_{k,l})_{central~det.}}
{(X  \sum_l Y_l \epsilon_{k,l}) _{central~det.} +
(X \sum_l Y_l\epsilon_{k,l})_{water}} ~,
\end{equation}
resulting in $Q_1 = 0.80 \pm 0.10$ and  $Q_2 = 0.94 \pm 0.07$.
In order to obtain the above values
a crude assumption $Y_1 = 2Y_2 = 2Y_3$, was used. 
It is important
to note that the uncertainties of $Q_k$ 
are strongly correlated with the
error in the efficiencies.

The effect of the neutron component
of hadronic cascades created outside the veto,
and coincident with the muon which created them, remains
to be determined. The ratio
of muons making three hits in the veto, thus
involving more than one particle, to the
prevalent case of the two hits was employed
as a  measure of the frequency of showers.
This ratio was $6.7 \pm 0.3 \%$ and $7.3 \pm 0.3 \%$
in the first and second runs, respectively, i.e., about
7\% of muons entering the veto were accompanied
by a shower.

To see that the `triple-veto-hit' events are 
really different from the standard throughgoing
muons with just two veto hits, one can form
ratios $R_k$
\begin{equation}
R_k = \frac{N_k^{shower}/N_k^{total}}
{N_{\mu}^{shower}/N_{\mu}^{total}} ~,
\end{equation}
which represent a quantitative measure of the
neutron content of the showers.
The measured values are $R_1 = 2.0 \pm 0.1$ and 
$R_2 = 4.4 \pm 0.2$, both significantly larger than unity.
Thus the shower events indeed are richer in neutrons,
particularly in the two neutron sample. 

Moreover, the double neutron events with three veto hits have distinctly
different time structure than the more common ones with just two
veto hits. This is shown in 
the bottom part of Fig. 2 for the interevent time,
i.e. the elapsed time between the first and second neutron captures.
The corresponding time constant $\tau_1$ for the three-veto-hit
events is significantly smaller
than for the two-veto-hit events, showing
that the shower events have a large
multineutron component. A similar effect is present
when the capture time of the first neutron is considered.
It is therefore likely  that the
discrepancies between data and simulation in the time dependence
of the interevent time interval, 
noted above, are the consequence of a multineutron ($k > 2$) component
in the two bank events. 

Since the shower events contain an unknown number of neutrons created
outside the detector volume, they are excluded from further 
consideration.


\section{Results}

In the first run, the observed numbers of neutron captures, 
corrected for the random background
and with the effect of showers subtracted, were $N_1 = 3916 \pm 66$, 
$N_2 = 828 \pm 29$,
while in the second run $N_1 = 3451 \pm 62$, $N_2 = 829 \pm 29$. The 
two runs, which were
separated in time by 10 months, give consistent results
when the differences in $\epsilon^{daq}$ is taken into
account, proving that the experiment
is stable. 

As a first step, the yields are analysed as in Ref. \cite{friends},
i.e., only the `diagonal' efficiencies $\epsilon_{k,k}$ are taken into
account, thus
\begin{equation}
Y_k^{simple} = N_k \cdot Q_k/(N_{\mu} \cdot X \cdot \epsilon_{k,k}
\cdot \epsilon_k^{daq}) ~.
\end{equation}
The resulting yields obtained by averaging the
two runs are:
\begin{eqnarray} 
Y_1^{simple} & = & (2.94 \pm 0.04 (stat) \pm 0.50 (syst)
)\times 10^{-5} 
\frac{{\rm neutrons}}{\mu \cdot {\rm g} \cdot {\rm cm}^{-2}} \\ \nonumber
Y_2^{simple} & = & (0.98 \pm 0.03 (stat) \pm 0.14 (syst)
) \times 10^{-5}  
\frac{{\rm neutrons}}{\mu \cdot {\rm g} \cdot {\rm cm}^{-2}} ~.
\end{eqnarray}

The exclusion of external showers, 
characterized by the three-veto-hit events,
has resulted in the reduction of $Y_1^{simple}$ by 14\%,
and of  $Y_2^{simple}$ by 45\%.
Thus, presumably due
to the presence of the sizable shield 
$X \simeq 380 ~ {\rm g/cm}^2$, the
effect of the external showers, while still
clearly present, was  reduced compared
to the findings of Ref. \cite{friends},
where (in the same units) 
$Y_1^{simple} = 4.3 (2.0) \times 10^{-5}$ and
$Y_2^{simple} = 1.6 (0.5) \times 10^{-5} $ without (with)
the correction for external showers.

However, the proper analysis should include the full efficiency
matrix, i.e., the possibility that two neutrons were initially
produced but only one neutron capture was recorded and vice versa,
as well as the possibility that three neutrons were originally
produced and only two or one neutron captures were observed.
As pointed out earlier, with only two measured capture rates,
$N_1$ and $N_2$, it is impossible to deduce all the relevant information
without further constraints. To avoid these difficulties,
and to make the comparison with other experiments easier,
the $total$ number of neutrons produced per muon was evaluated.
This quantity, $Y_{tot}$, defined earlier in Eq. (2),
has the further advantage that it is
essentially independent of the ratio $Y_3/Y_2$ for three to two
neutron production, as demonstrated in Table I.

Thus, the final result of the present measurement can be expressed
as 
\begin{equation}
Y_{tot} = (3.60  \pm 0.09 \pm 0.31 )
\times 10^{-5} 
\frac{{\rm neutrons}}{\mu \cdot {\rm g} \cdot {\rm cm}^{-2}}~,
\end{equation}
where the systematic error is  an estimate based
on the spread of values in Table I added in quadrature to the
spread of the evaluated efficiencies.

All available data on neutron yields are collected 
and compared in Fig. 3.
(It is assumed that the other measurements 
(\cite{Bezrukov,Enikeev,Aglietta,Aglietta2}) also
are really $Y_{tot}$ measurements.)
For the measurements of Ref. \cite{friends} both results, 
with and without the correction for external showers,
are shown.
In some of the other measurement the shower contribution
was excluded, but it is not clear how well this has been
done, since at least the results at 25 and 316 mwe
were obtained with relatively small and unshielded detectors.

At shallow depth there are now three measurements, with 
essentially consistent results. Clearly, still   
better and more complete measurements are desirable, in which the 
full neutron multiplicity and energy spectra are
determined and the
neutrons produced externally are reliably identified.

\begin{table}[htb]
\caption{ Values of the neutron yields $Y_{tot}$, $Y_1$,
$Y_2$, and $Y_3$ in units of $10^{-5}
\frac{{\rm neutrons}}{\mu \cdot {\rm g} \cdot {\rm cm}^{-2}}$
for different assumed ratios $Y_3/Y_2$.}
\begin{center}
\begin{tabular}{|r|c|c|c|c|} 
Assumed $Y_3/Y_2$ & $Y_{tot}$ & $Y_1$ & $Y_2$ & $Y_3$ \\ \hline
0. & 3.54 & 2.30 & 0.62 & 0.0 \\
0.5 & 3.60 & 2.48 & 0.32 & 0.16 \\
1. & 3.62 & 2.54 & 0.22 & 0.22 \\
2. & 3.64 & 2.59 & 0.13 & 0.26  
\label{tab:Ytot}
\end{tabular}
\end{center}
\end{table}


\subsection{Acknowledgment}

We would like to thank the Arizona Public Service Company for
the generous hospitality provided at the Palo Verde plant.
The important contributions of M.~Chen, R.~Hertenberger, K.~Lou, and
N.~Mascarenhas in the early stages of this project are gratefully
acknowledged.
We also acknowledge the generous financial help from
the University of Alabama, Arizona State University,
California Institute of Technology, and
Stanford University. Finally, our gratitude goes to CERN, DESY, FNAL,
LANL, LLNL, SLAC, and TJNAF who at different times provided us with
parts and equipment needed for the experiment.
This project was supported in part by the USDoE.


\begin{figure}
\begin{center}
\mbox{\psfig{figure=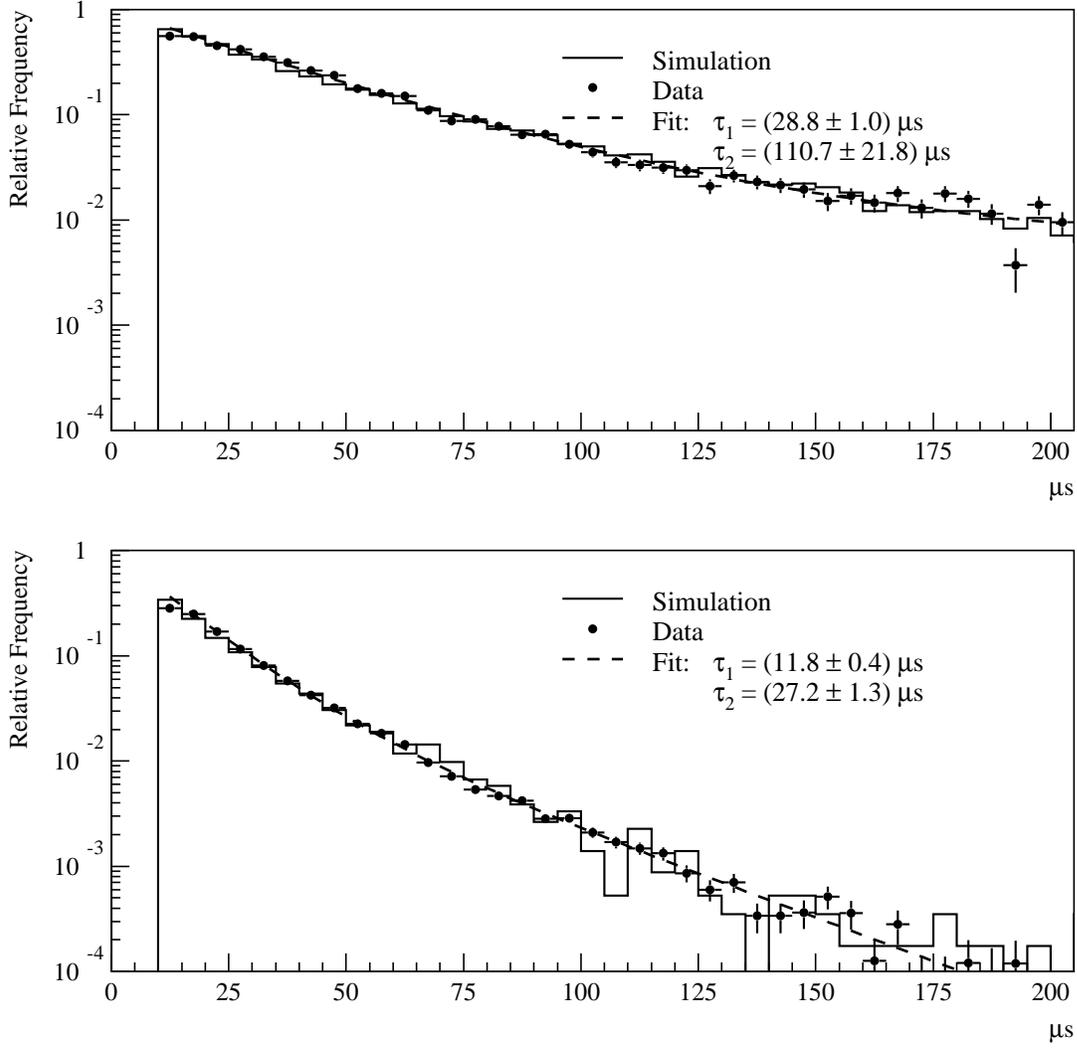,width=16cm}}
\caption{\leftskip=3pc\rightskip=3pc\noindent
Top: The distribution of time between  muon hits and the
single-neutron capture events, fitted to the 
exponential function, Eq. (3).
Bottom: The distribution of time between  muon hits and the
first neutron capture events for the
double-neutron events with an exponential fit, Eq. (3).
In both panels the MC simulation is shown as a histogram,
and the fitted time constants $\tau_1$ and $\tau_2$ are displayed.}
\label{ti_veto}
\end{center}
\end{figure}

\begin{figure}
\begin{center}
\mbox{\psfig{figure=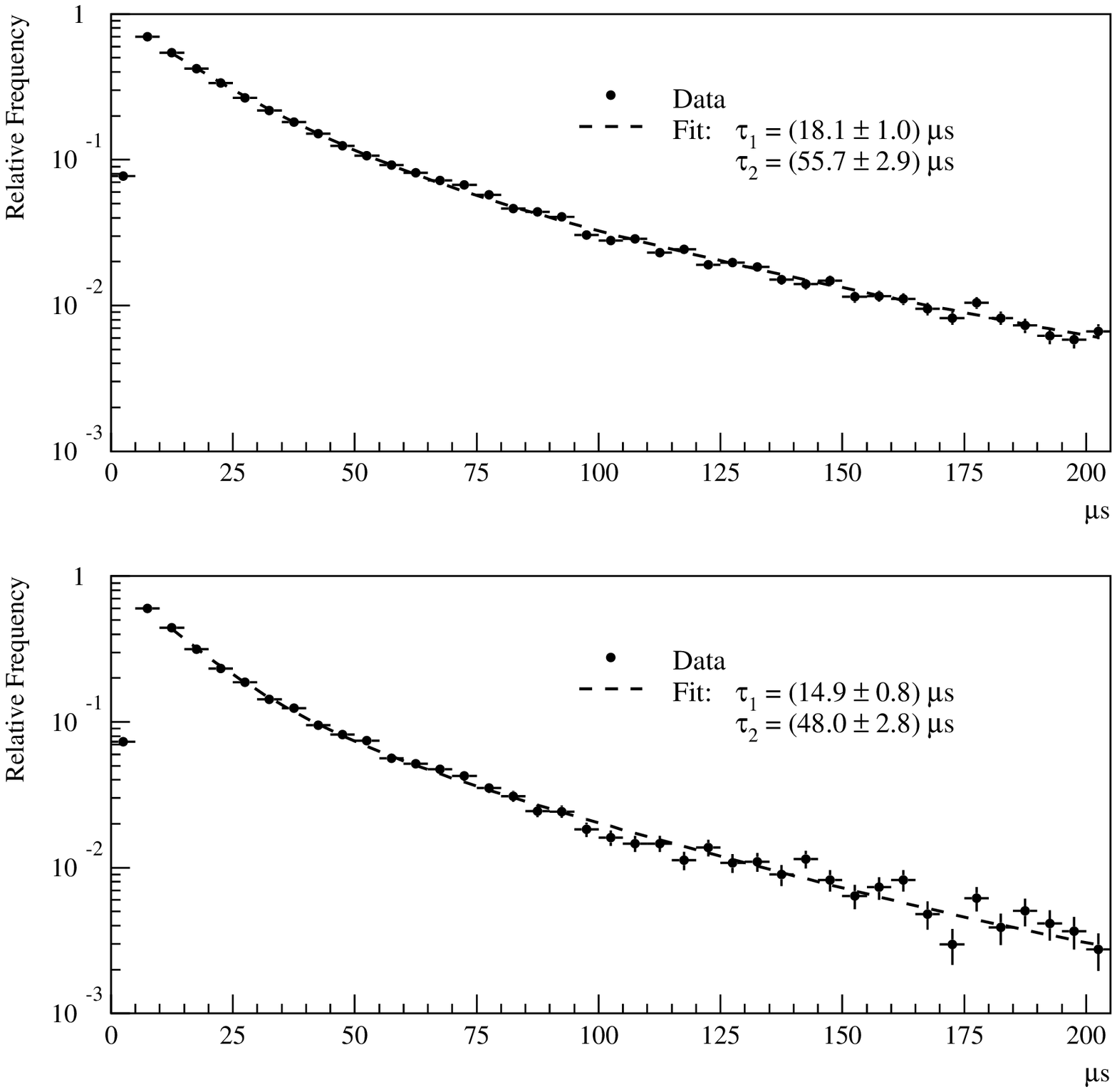,width=16cm}}
\caption{\leftskip=3pc\rightskip=3pc\noindent
Top: The distribution of the elapsed time between the two
capture events with the two-veto-hit pattern for the
double-neutron events with an exponential fit, Eq. (3).
Bottom: The distribution of the elapsed time between the two
capture events for the subset of the
double-neutron events with three veto hits. 
In both panels 
the fitted time constants $\tau_1$ and $\tau_2$ are displayed.}
\label{ti_inter}
\end{center}
\end{figure}

\begin{figure}
\epsfysize=5.0in \epsfbox{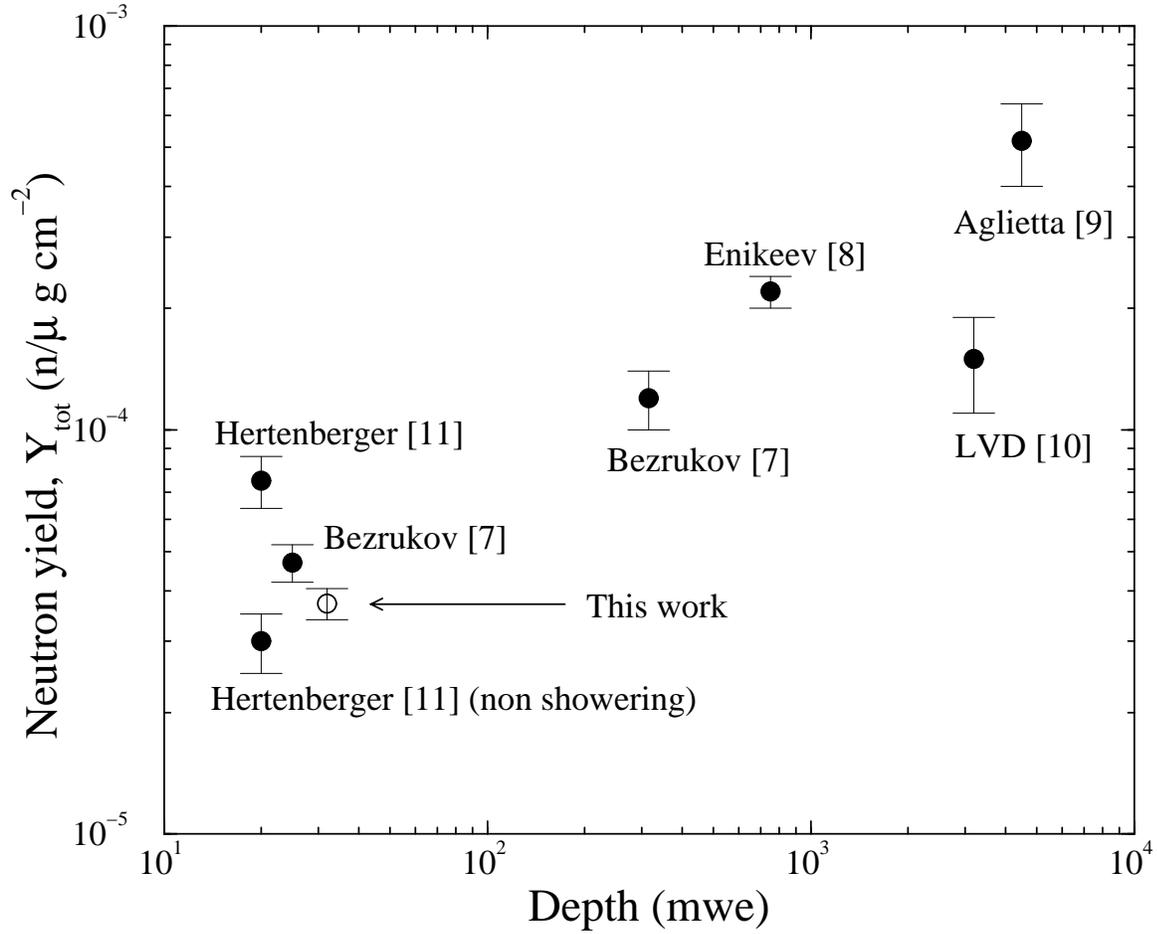}
\caption{\protect{\leftskip=3pc\rightskip=3pc\noindent
Summary of available data on the total neutron yield 
$Y_{tot}$ versus depth.
The points are labelled by the corresponding first author.
For Ref. [11] two results are shown, with and without
the effect of external showers.Present
work is denoted by an empty circle.}}
\label{fig:summary}
\end{figure}


\end{document}